\documentclass[conference]{IEEEtran}
\IEEEoverridecommandlockouts
\usepackage{cite}
\usepackage{amsmath,amssymb,amsfonts}
\usepackage{algorithmic}
\usepackage{graphicx}
\usepackage{multirow}%
\usepackage{listings}%
\usepackage{caption}
\usepackage{subcaption}
\usepackage{float} 
\usepackage{wrapfig}
\usepackage{array}
\usepackage{textcomp}
\usepackage{pdfpages}
\usepackage{flushend}

\def\BibTeX{{\rm B\kern-.05em{\sc i\kern-.025em b}\kern-.08em
    T\kern-.1667em\lower.7ex\hbox{E}\kern-.125emX}}
\begin{document}

\title{Gamified Virtual Reality Exposure Therapy for Mysophobia: Evaluating the Efficacy of a Simulated Sneeze Intervention\\
}

\author{\IEEEauthorblockN{1\textsuperscript{st} Md Mosharaf Hossan}
\IEEEauthorblockA{\textit{Department of Computer Science} \\
\textit{Idaho State University}\\
Pocatello, United States \\
mdmosharafhossan@isu.edu}
\and
\IEEEauthorblockN{2\textsuperscript{nd} Rifat Ara Tasnim}
\IEEEauthorblockA{\textit{Department of Computer Science} \\
\textit{Idaho State University}\\
Pocatello, United States \\
rifataratasnim@isu.edu}
\and
\IEEEauthorblockN{3\textsuperscript{rd} Farjana Z Eishita}
\IEEEauthorblockA{\textit{Department of Computer Science} \\
\textit{Idaho State University}\\
Pocatello, United States \\
farjanaeishita@isu.edu}
}

\maketitle

\begin{abstract}
Mysophobia, or the fear of germs, is a prevalent anxiety disorder that significantly impacts daily life. This study investigates the potential of a gamified virtual reality (VR) intervention to simulate contamination-related scenarios and assess their emotional and psychological effects. A VR game based sneeze simulation was developed to evaluate its influence on participants' emotional states. Seven participants completed two versions of the game: a baseline version and an experimental version featuring the sneeze simulation. Emotional responses were measured using the Positive and Negative Affect Schedule (PANAS) and State-Trait Anxiety Inventory - State (STAI-S) questionnaires. The results revealed slight increases in negative affect and anxiety levels during the sneeze simulation. Also, a reduction in positive affect was revealed. However, these differences were not statistically significant (p \textgreater 0.05). This is likely due to small sample sizes, a lack of grossness in the simulation, or participants not being clinically mysophobes. This exploratory study highlights the potential of VR-based interventions for understanding and addressing contamination-related anxieties. It provides a foundation for future research with larger and more diverse participant pools.
\end{abstract}


\section{Introduction}

Mysophobia, commonly known as the fear of germs, is a significant anxiety disorder that manifests as an excessive preoccupation with cleanliness and avoidance of contamination. This condition can severely impact an individual's daily life and lead to behaviors such as frequent handwashing, avoidance of public spaces, and heightened anxiety in situations involving perceived exposure to germs. Traditional therapeutic approaches such as cognitive-behavioral therapy (CBT) and exposure therapy have been effective in addressing mysophobia. However, the risks associated with in-person exposure therapy, such as actual germ exposure, highlight the need for innovative and safer alternatives.

Virtual Reality (VR) technology has emerged as a transformative tool in mental health interventions. It offers controlled and immersive environments for exposure therapy. By recreating anxiety-inducing scenarios in a safe and customizable manner, VR enables individuals to confront and manage their fears. Gamification of such scenarios can enhance engagement. Studies on gamified VR applications have demonstrated their efficacy in treating various phobias and anxiety disorders. For instance, Lindner et al. \cite{lindner2020gamified} \cite{lindner2020experiences}  highlighted the effectiveness of gamified VR exposure therapy (VRET) for spider phobia. They emphasized how gamification elements such as task progression and narrative structure transformed traditional therapy into an engaging and immersive experience. Similarly, Miloff et al. \cite{miloff2016single} established the potential of gamified VRET as a viable alternative to traditional exposure therapy. They demonstrated comparable outcomes with greater accessibility and logistical ease.

Recent advancements in adaptive VR systems have expanded their applicability to broader psychological conditions. Davies et al. \cite{davies2024immersive} demonstrated the value of integrating behavioral tracking into VR games to tailor gameplay dynamically based on participants' physiological and emotional states. This adaptive approach enhances therapeutic outcomes by aligning the intervention with individual needs in real-time. Furthermore, Kahlon et al. \cite{kahlon2023gamified} validated the efficacy of gamified VRET in reducing public speaking anxiety among adolescents.

In the context of mysophobia, Yuan et al. \cite{yuan2018application} modeled obsessive-compulsive disorder (OCD) in a conscious system. They focused on symptoms such as compulsive handwashing. This work underscores the relevance of simulating contamination-related behaviors for targeted interventions. Meanwhile, Craske \cite{craske2015optimizing} explored strategies to optimize exposure therapy with inhibitory learning to disconfirm threat expectations. Eishita et al. \cite{eishita2023gamified} expanded on these approaches by proposing a prototype named Militant of the Maze Generation 2 (MoMG), a gamified handheld digital platform for treating OCD-related cleanliness compulsions. By incorporating virtual contamination scenarios and real-time heart rate variability (HRV) monitoring, MoMG trains users to manage their anxiety during simulated exposure tasks.

Building on these strategies, this study aims to assess the efficacy of a VR-based gamified intervention designed to simulate exposure to a sneeze, a common trigger for individuals with mysophobia. Participants engage in a shooting game under two conditions: a baseline scenario without contamination stimuli and a modified scenario featuring a simulated sneeze. Gamification elements like scoring and time limits make the experience engaging and help collect reliable data.

We explore the following research questions:
 
\begin{enumerate}
    
    \item What are the perceived experience from digital contamination stimuli within a VR environment? 

    \item Can exposure to digital contamination stimuli in VR reliably induce measurable anxiety responses?


\end{enumerate}

The study evaluates emotional responses using validated questionnaires such as the Positive and Negative Affect Schedule (PANAS) and the State-Trait Anxiety Inventory - State (STAI-S) to measure changes in emotional state before and after exposure to the sneeze simulation. By leveraging insights from prior research on gamified and adaptive VR systems, this study seeks to contribute to the development of scalable and effective therapeutic tools for mysophobia and related contamination anxieties. 


\section{ Literature Review}

The application of gamified and automated Virtual Reality Exposure Therapy (VRET) systems has demonstrated promising results in treating specific phobias and anxiety disorders. Lindner et al. \cite{lindner2020gamified} showcased the efficacy of a gamified VRET system for spider phobia. Significant symptom reduction was observed through a single-session, self-guided intervention under simulated real-world conditions. This highlights the scalability and effectiveness of combining gamification and automation in therapeutic settings. Additionally, Lindner et al. \cite{lindner2020experiences} emphasized the role of gamification elements such as task progression and narrative structure in enhancing user engagement and satisfaction.
These findings show the value of using gamification to boost engagement and improve results.

Miloff et al. \cite{miloff2016single} further expanded the scope of gamified VRET by comparing its efficacy to traditional one-session exposure therapy (OST) for spider phobia. Their randomized controlled trial demonstrated that gamified VRET was non-inferior to OST while providing greater accessibility and reducing logistical barriers. These findings validate the potential of gamified VR applications as viable alternatives to traditional therapy.

Beyond specific phobias, gamified VR interventions have been explored for broader psychological conditions. Davies et al. \cite{davies2024immersive} developed a VR first-person shooter game with behavioral tracking sensors to treat hypervigilance. This approach exemplifies how VR systems can be personalized to promote emotional and cognitive regulation. Similarly, Hossan et al. \cite{hossan2024adaptive} discussed different personalization techniques to adapt games to player preferences.

Kahlon et al. \cite{kahlon2023gamified} applied gamified VRET to reduce public speaking anxiety in adolescents. They demonstrated significant symptom reduction through an engaging and self-guided intervention. The inclusion of gamification elements such as progress tracking and feedback improved user engagement and supported consistent adherence to the therapy. These studies underscore the value of adaptive and immersive VR systems in addressing diverse psychological challenges.

In the context of mysophobia, Yuan et al. \cite{yuan2018application} proposed a conscious system to model obsessive-compulsive disorder (OCD), specifically targeting compulsive behaviors like excessive handwashing. By integrating cognitive-behavioral therapy (CBT) principles into an artificial intelligence based framework, the study highlights the potential for technology-driven mental health interventions. Meanwhile, Yameen and Qadir \cite{qadir2019questionnaire} investigated the physiological underpinnings of mysophobia and revealed no significant correlation between blood oxygen levels and germ-related anxiety. This finding reinforced the need for psychological, rather than physiological, approaches in designing interventions for mysophobia.

To optimize exposure-based therapies, Craske \cite{craske2015optimizing} explored strategies rooted in inhibitory learning and regulation models. Techniques such as deepened extinction and occasional reinforced extinction were shown to enhance the effectiveness of exposure therapy by disconfirming threat expectations. These methodologies offer a foundation for designing exposure based interventions that maximize therapeutic outcomes. Finally, Page and Coxon \cite{page2016virtual} reviewed the effectiveness of VRET for anxiety disorders. They highlighted the potential of VR environments to provide controlled, customizable exposure scenarios. While they identified methodological limitations such as small sample sizes and lack of control groups, the review reinforced the validity of VR as a therapeutic tool.

Eishita et al. \cite{eishita2023gamified} proposed Militant of the Maze Generation 2 (MoMG), a gamified handheld digital platform for treating OCD-related cleanliness compulsions. By incorporating virtual contamination scenarios, animated germ splashes, and real-time heart rate variability (HRV) monitoring, their system aim to train users in emotional regulation during exposure tasks. 

Inspired by these studies, we designed our VR-based intervention for mysophobia to test the efficacy of a simulated sneeze scenario. By integrating validated methodologies from these prior works, we aimed to establish a robust framework for evaluating the role of VR in treating mysophobia, advancing the understanding of its therapeutic potential.


\section{Methodology}


\subsection{Participants}

The study involved 7 participants (4 males and 3 females) who voluntarily took part in the experiment. All participants were briefed about the study procedures, and a signed consent letter was obtained prior to their participation. Inclusion criteria required participants to have no prior history of diagnosed mysophobia. We initially conducted the experiment on non-mysophobic individuals to evaluate the game's impact on emotional states in a controlled manner. 


\subsection{ Materials and Setup}

A virtual reality (VR) game was developed to simulate a controlled contamination related scenario. The game was designed to include two versions: a baseline condition without contamination stimuli and an experimental condition incorporating a digital sneeze simulation. The VR system consisted of a oculus meta quest 3 head mounted display with motion tracking to ensure an immersive and interactive experience. The game environment included a virtual room with tables, non-playable characters (NPCs), and target objects.

The Positive and Negative Affect Schedule (PANAS) and the State-Trait Anxiety Inventory - State (STAI-S) questionnaires were used to measure participants' emotional states before and after exposure to the scenarios.

Each participant experienced two versions of the VR game sequentially:

\subsection{Baseline Version}
\begin{figure}[H] 
    \centering
    \includegraphics[width=0.4\textwidth]{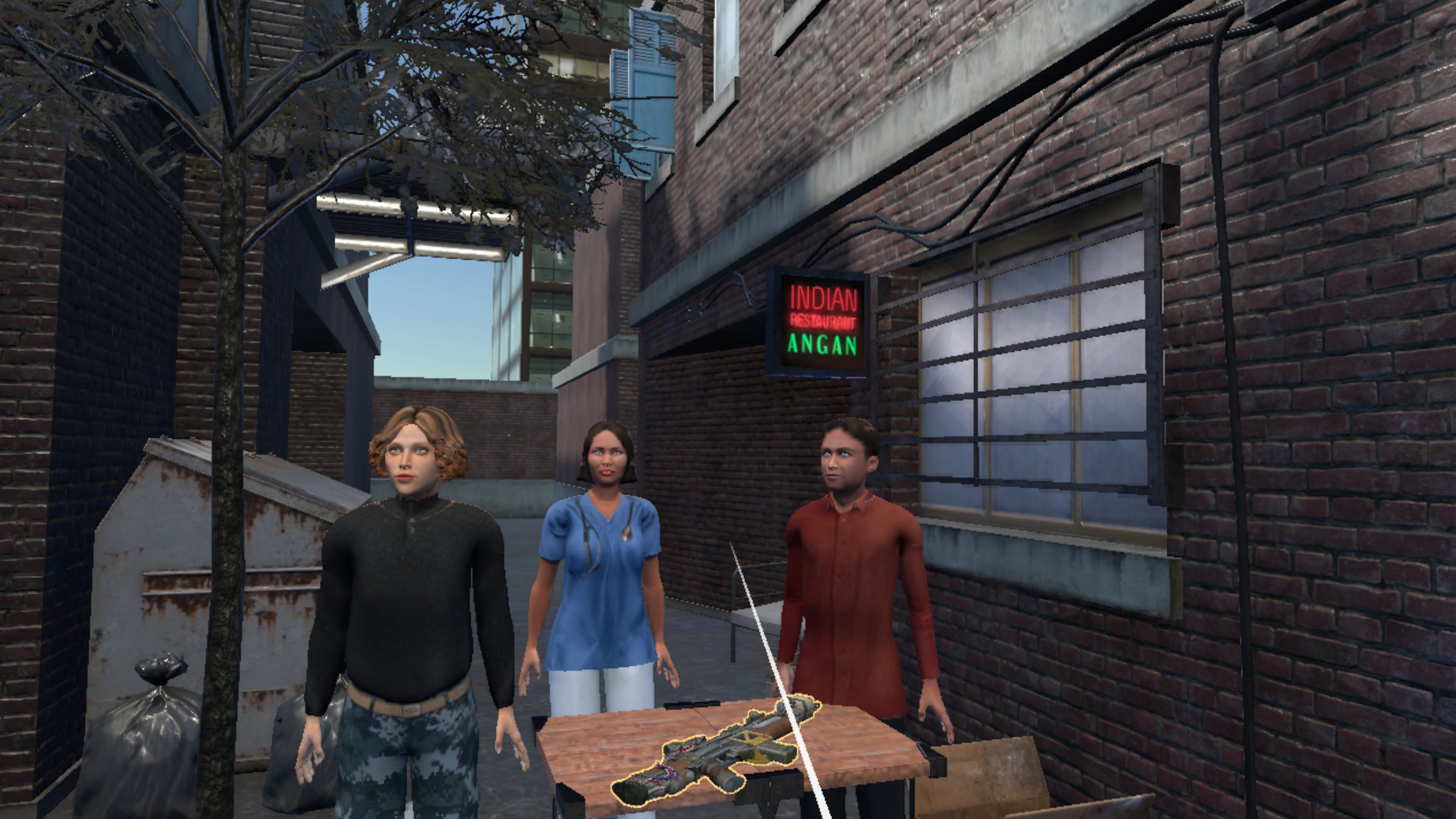} 
    \caption{Gun pickup of baseline version}
    \label{fig: Gun pickup of baseline version} 
\end{figure}

\begin{figure}[H] 
    \centering
    \includegraphics[width=0.4\textwidth]{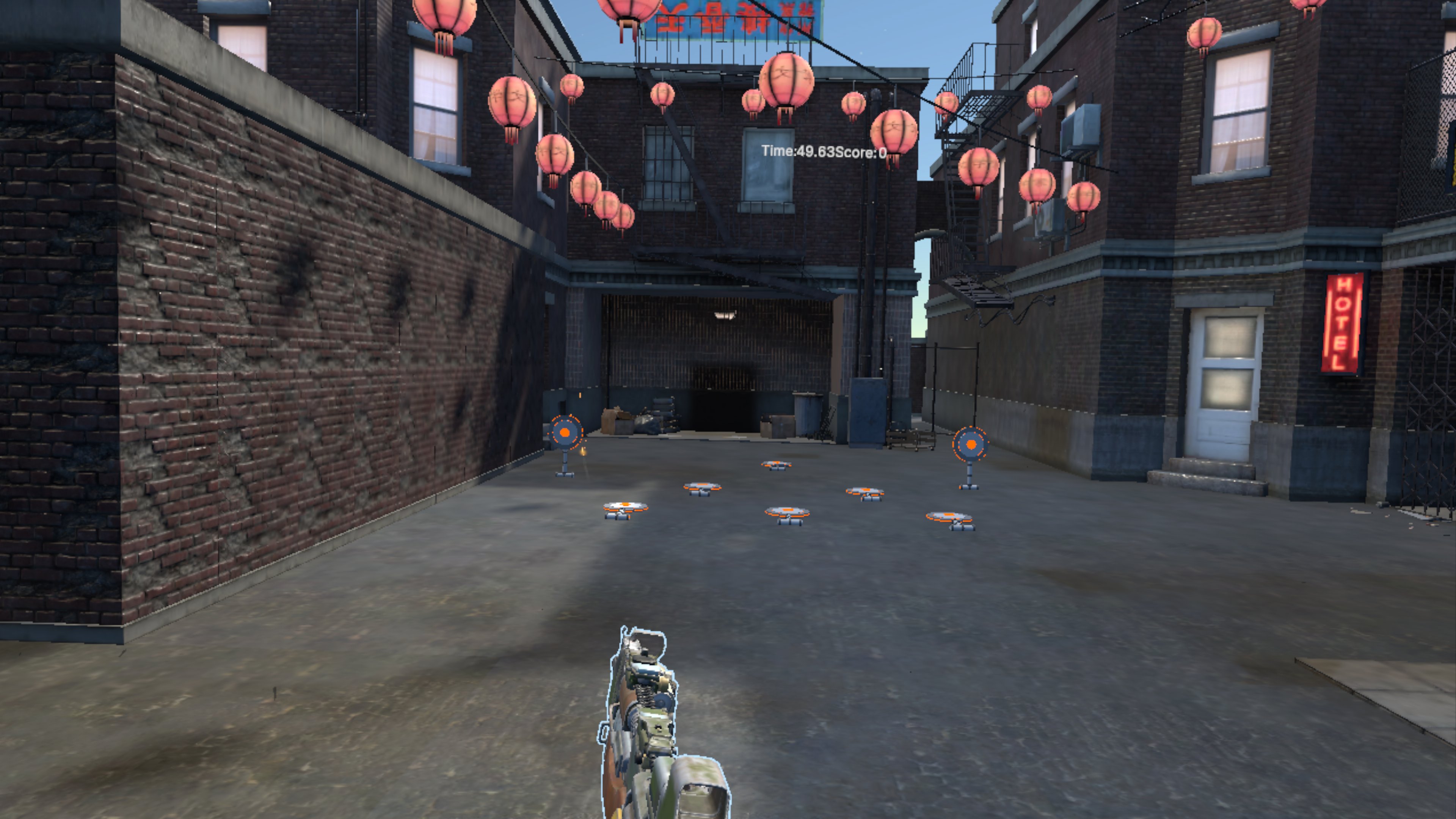} 
    \caption{Gameplay of baseline version}
    \label{fig: Gameplay of baseline version} 
\end{figure}

   Participants were instructed to navigate the virtual environment and complete a series of tasks. First, as shown in Figure \ref{fig: Gun pickup of baseline version} they walked to a table to pick up a rifle, which was placed near three NPCs. After acquiring the rifle, participants proceeded to another table and initiated the game by pressing a start button. After the start of the game moving targets in random positions appeared in front of the player as seen in Figure \ref{fig: Gameplay of baseline version}.
   Participants were tasked with shooting the targets within a 1 minute time frame. Score were calculated based on accuracy with bullseye shots awarding the highest score. A timer displayed the remaining time and a score indicator tracked the participant's performance. Upon completion, participants reported their scores to the experimenter and completed the PANAS and STAI-S questionnaires.

\subsection{Experimental Version}
\begin{figure}[H] 
    \centering
    \includegraphics[width=0.4\textwidth]{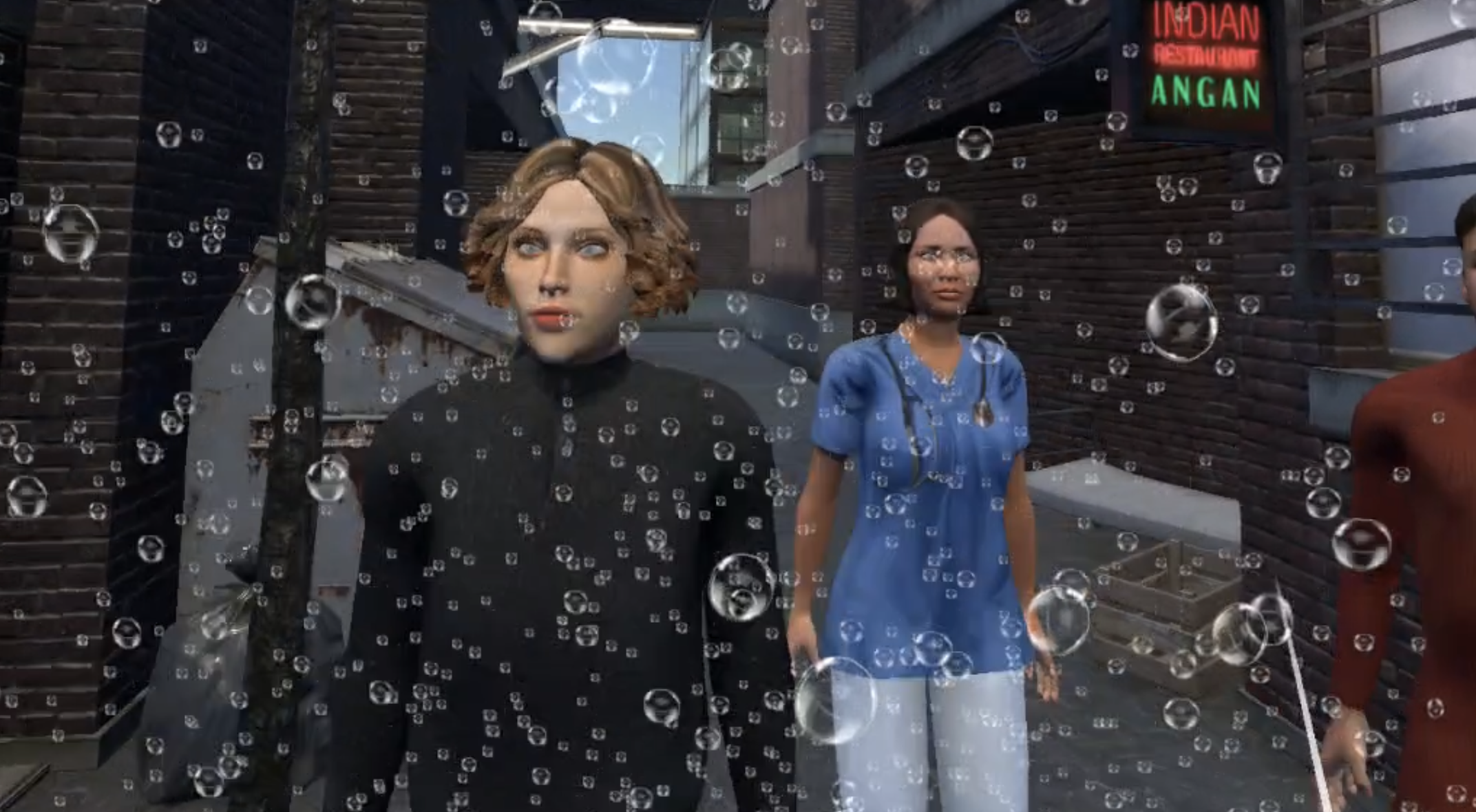} 
    \caption{Sneeze droplets on face in experimental version}
    \label{fig: Sneeze droplets on face in experimental version} 
\end{figure}

\begin{figure}[H] 
    \centering
    \includegraphics[width=0.4\textwidth]{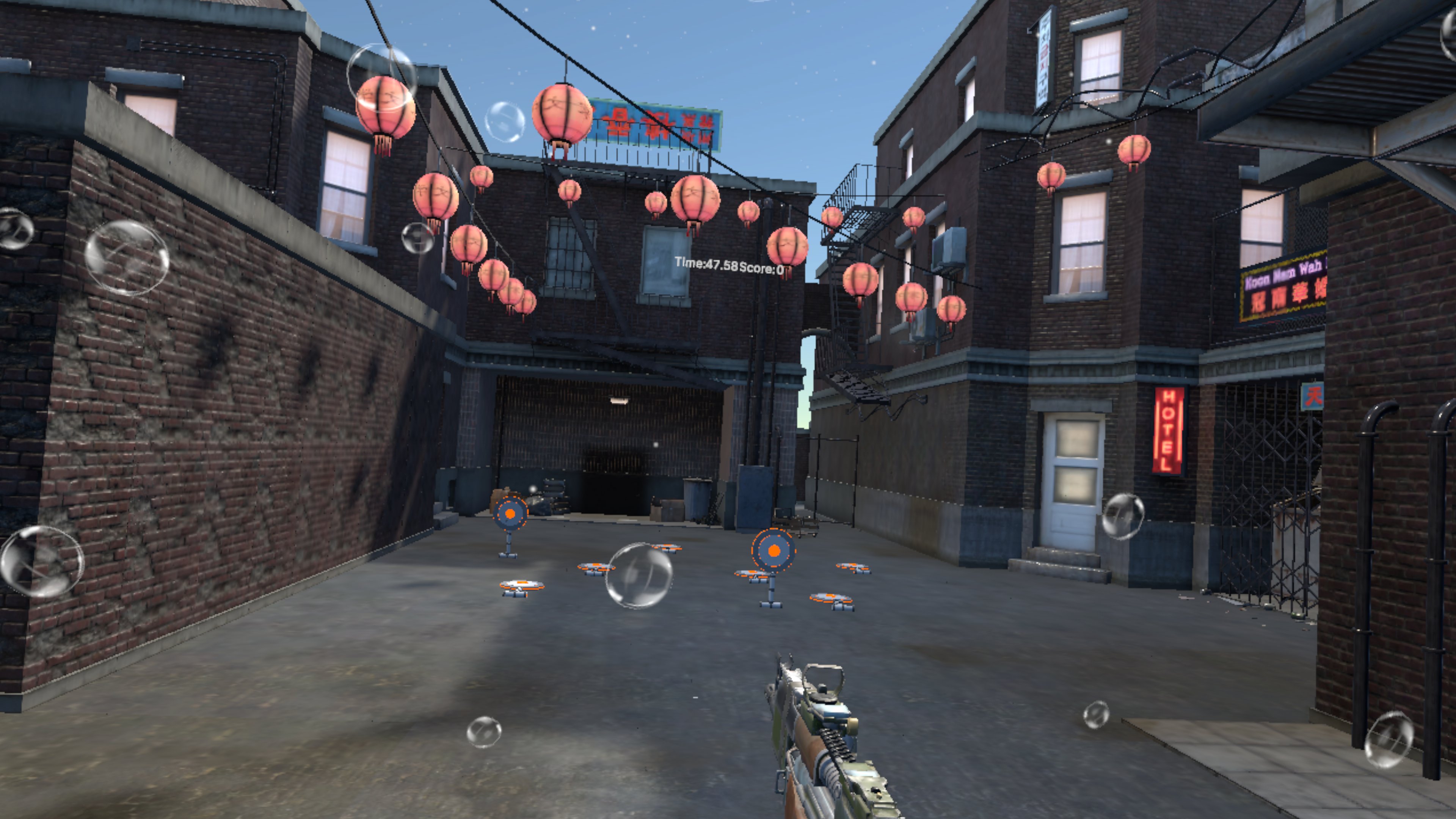} 
    \caption{Gameplay of experimental version}
    \label{fig: Gameplay of experimental version} 
\end{figure}

The second version of the game included a contamination stimulus. As participants approached the table to pick up the rifle, two of the three NPCs standing near the table sneezed at the player as shown in Figure \ref{fig: Sneeze droplets on face in experimental version}. A gross sneeze sound was added for both male and female NPCs to make it more realistic. These NPCs were chosen so that the player would get sneezed on regardless of the direction they approached the table. After the sneezes, digital droplets appeared on the participant’s headset . This created a realistic contamination effect. Participants then played the target shooting game for one minute under the same conditions as the baseline version while having few of the droplets still on the headset as shown in Figure \ref{fig: Gameplay of experimental version}. After finishing, they reported their scores and completed the PANAS and STAI-S questionnaires.

\subsection{Questionnaires}

\subsubsection{Positive and Negative Affect Schedule (PANAS)}
The Positive and Negative Affect Schedule (PANAS) was created by Watson et al. \cite{watson1988development} in 1988 for self-report measurement of positive and
negative affect of an individual. It consists of 20 scales, 10
for each affect. It include Interested, Distressed, Excited, Upset, Strong, Guilty, Scared, Hostile, Enthusiastic, Proud, Irritable, Alert, Ashamed, Inspired, Nervous, Determined, Attentive, Jittery, Active, and Afraid. Each item contains a 5-point scale of 1 to 5 where 1 represents very slightly or not at all and 4 stands for extremely. Positive and negative affects are
calculated by adding the scores of corresponding items.

\subsubsection{State Trait Anxiety Inventory - State (STAI-S)}

Spielberger et al. \cite{spielberger1971state} developed the State Trait Anxiety Inventory (STAI) to assess a subject's perceived anxiety after a stressful situation. State anxiety refers to a temporary feeling of anxiety triggered by a specific situation, while trait anxiety is a more stable personality trait that indicates a general tendency to experience anxiety in stressful situations. For this experiment we only looked at the State anxiety of participants.

\subsection{Data Collection}

The primary data collected included:
1. Scores from the target-shooting game: Recorded in both baseline and experimental conditions to assess the impact of the sneeze simulation on task performance.
2. PANAS and STAI-S questionnaire responses: Administered after each game version to evaluate changes in participants' emotional states. The collected data were analyzed to compare participants' emotional responses and performance across the two conditions.

\section*{Experiment}

Total 7 university students aged between 20 to 30 were recruited through email invitation. A controlled lab experiment was conducted
with a consent letter being signed by each participant. They were prohibited from distractions during gameplay. Before gamplay, participants were given 2 minute to familiarize themselves with the environment and controls.
Participants played
1 sessions of each version and completed questionnaires afterward. The In dependent Variable is the Game version with 2 conditions - Version 1 and Version 2. The Dependant Variables include PANAS
and STAI-S. players’ scores are the considered to the performance
metrics. The experiment was designed to be conducted in a
within subject fashion. The sessions were ordered using the Latin Square Method \cite{keedwell2015latin}.

\section*{Result}

The results from the study were analyzed using the Positive and Negative Affect Schedule (PANAS) and State-Trait Anxiety Inventory - State (STAI-S) questionnaires to evaluate participants' emotional responses to the virtual sneeze simulation. A paired t-test was used to assess differences between the baseline (plain) condition and the sneeze condition. The findings are detailed below:

\subsection{ PANAS Positive Affect (PA) }
\begin{figure}[H] 
    \centering
    \includegraphics[width=0.4\textwidth]{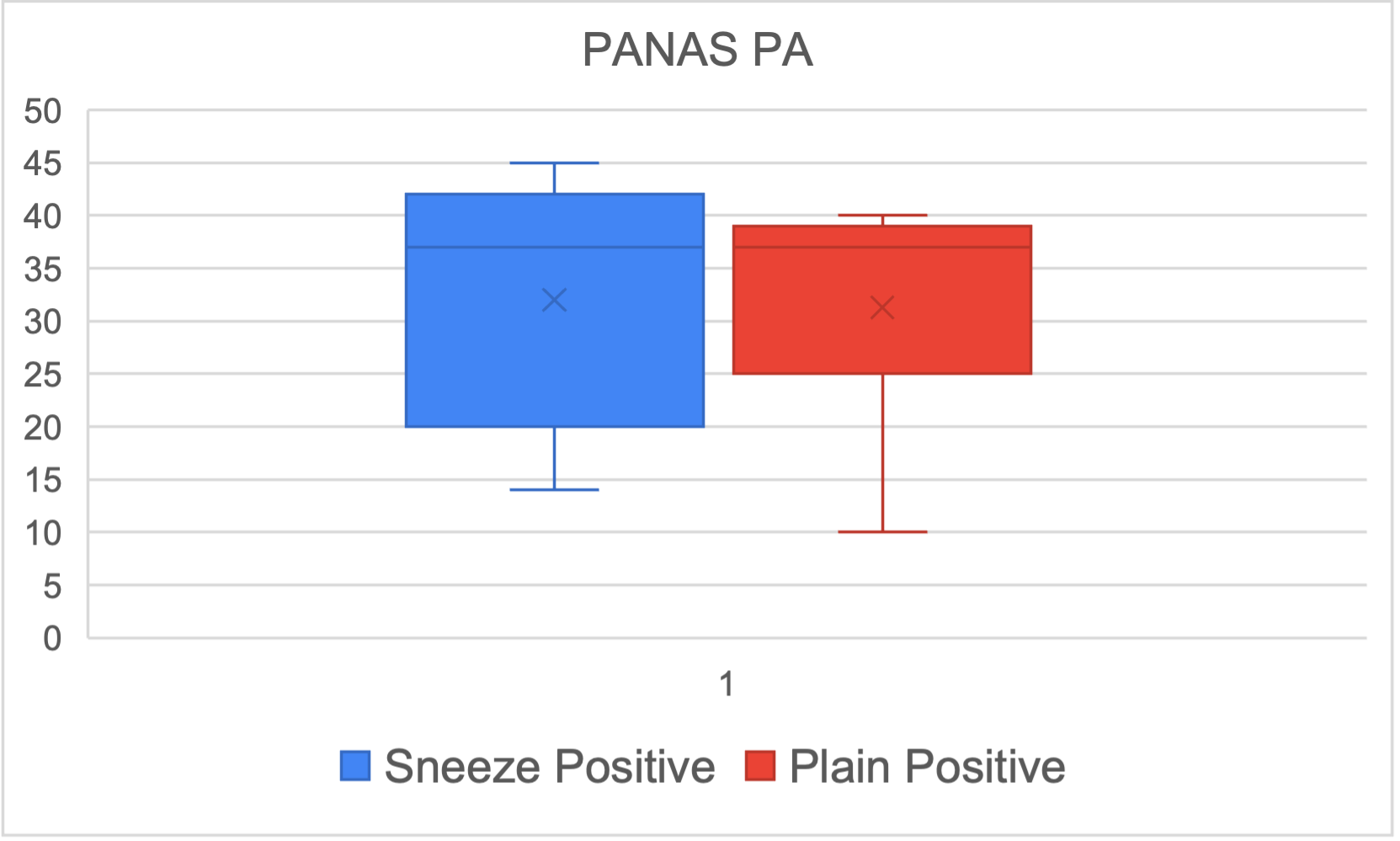} 
    \caption{Positive affect}
    \label{fig: Positive affect} 
\end{figure}

For Positive Affect (PA),the p-value (p = 0.34) indicates that the reduction in positive affect in the sneeze condition was not statistically significant. The boxplot (Figure \ref{fig: Positive affect}) highlights the broader range of scores observed in the sneeze condition.

 
\subsection{ PANAS Negative Affect (NA) }
\begin{figure}[H] 
    \centering
    \includegraphics[width=0.4\textwidth]{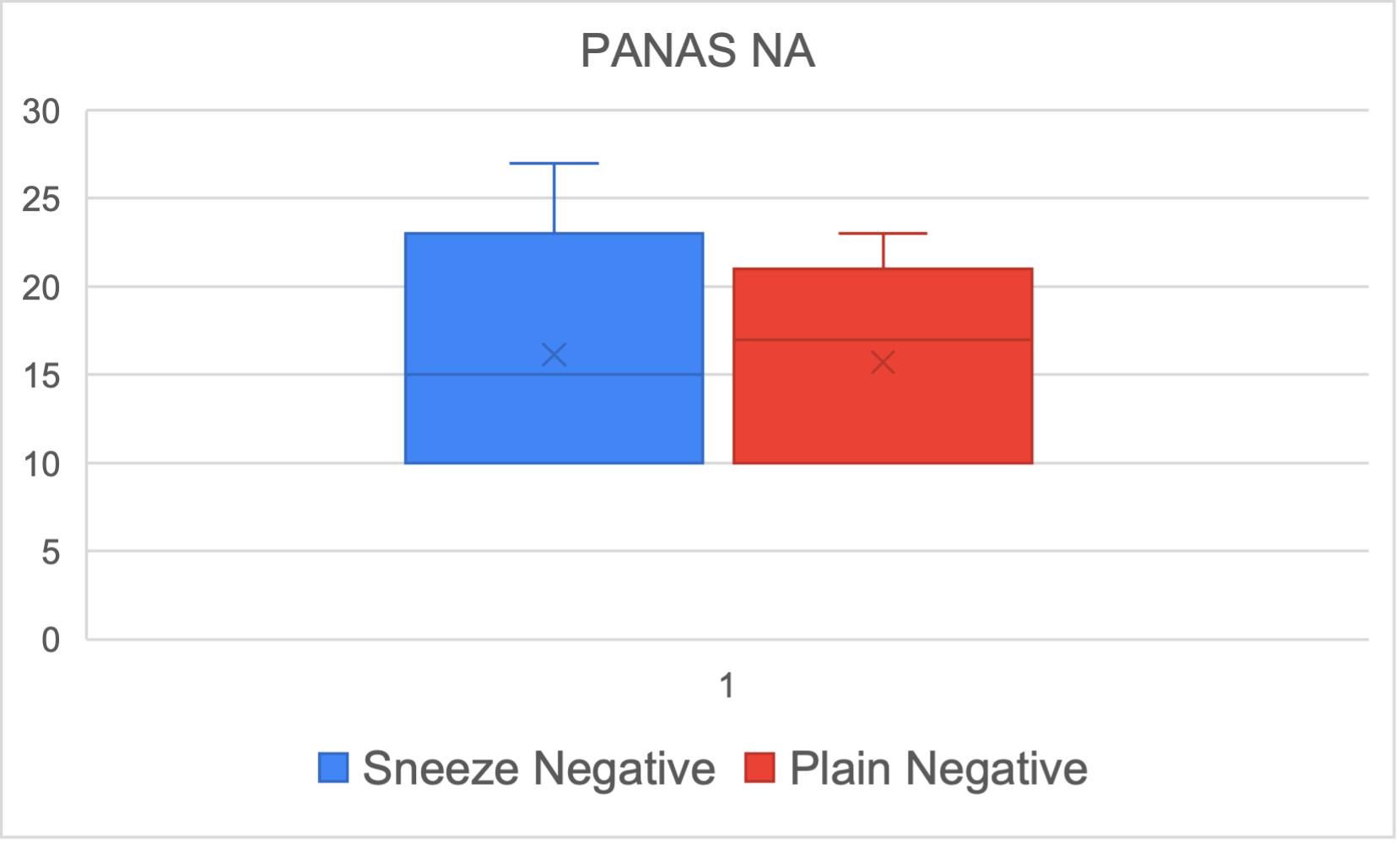} 
    \caption{Negative affect}
    \label{fig: Negative affect} 
\end{figure}

The average Negative Affect (NA) score for the sneeze condition was slightly higher compared to the plain condition. However, the difference was not statistically significant (p = 0.44). The boxplot (Figure \ref{fig: Negative affect}) shows that the sneeze condition induced marginally higher variability in negative affect. It suggests a potential emotional impact of the sneeze simulation.

\subsection{STAI-S}
\begin{figure}[H] 
    \centering
    \includegraphics[width=0.4\textwidth]{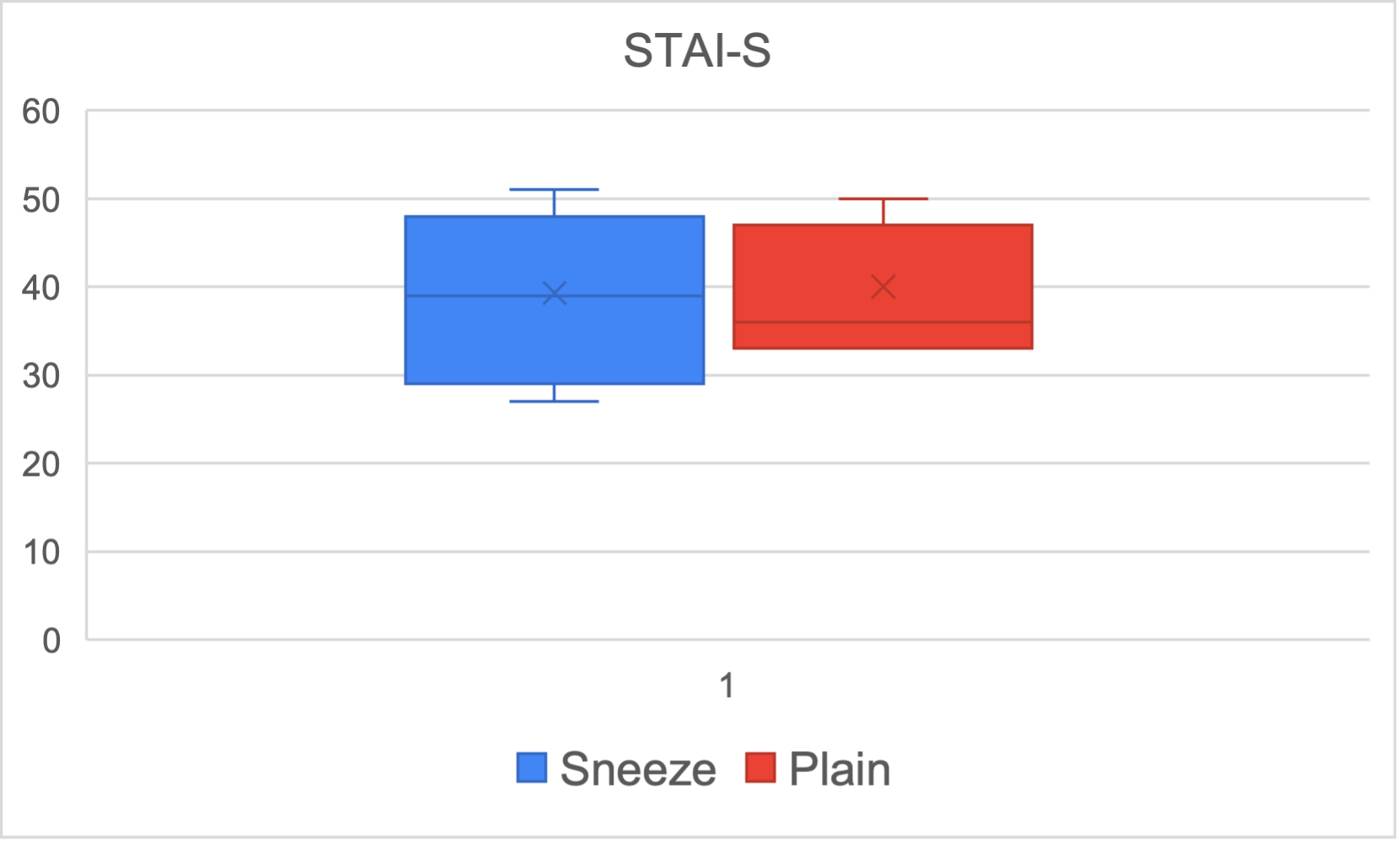} 
    \caption{State Trait anxiety inventory - State}
    \label{fig: State Trait anxiety inventory - State} 
\end{figure}

The State Anxiety (STAI-S) scores were analyzed to assess participants' anxiety levels. The average scores were higher in the sneeze condition compared to the plain condition. However, the difference was again not statistically significant (p = 0.33). Figure \ref{fig: State Trait anxiety inventory - State} shows that participants experienced some variability in anxiety levels. This may reflect differences in individual sensitivity to the sneeze simulation.

\section*{Discussion}

The findings of this study suggest that VR based gamified interventions can provide an innovative approach to assessing and potentially mitigating contamination related anxieties, such as those associated with mysophobia. The immersive nature of VR allowed for a controlled environment to simulate a realistic yet safe exposure to a triggering scenario—sneezing. This approach engages participants and enables researchers to collect detailed data on their emotional and behavioral responses. The sneeze simulation had a measurable impact as shown in the PANAS and STAI-S questionnaire results. This highlights the potential of VR technology as a useful tool for studying and treating specific phobias.

The statistical analysis suggests that while there were observable trends in the emotional and anxiety responses to the virtual sneeze simulation, these differences were not statistically significant across the small sample size (n = 7). Participants not being mysophobic could also be a significant factor. However, during experiment no adverse effect e.g. seizures were reported and the sneeze version elicited more negative emotions and varied anxiety response. These outcomes pave the way to conduct experiment safely on mysophobes. Future research with mysophobic participant pool and additional variables could provide more robust insights into the emotional and psychological impacts of contamination related VR simulations. Additionally, the findings pave the way for integrating biofeedback and real-time emotional regulation mechanisms into VR games. This approach advances research on mysophobia and establishes a foundation for developing interventions targeting a wide range of anxiety disorders in an engaging, scalable, and cost-effective manner.


\paragraph{Threats to Validity}
The small sample size of 7 non-mysophobic participants (4 males and 3 females) limits the generalizability of the results. A larger and more focused participant pool would provide more robust and representative data. Additionally, the use of self-reported measures, such as PANAS and STAI-S, may introduce bias due to participants' subjective interpretation of their emotional states. Lastly, the game design and the VR environment may have affected the results. Changes in game mechanics or the realism of the sneeze could lead to different outcomes.

\paragraph{Future Work}
Future research should address these limitations and explore broader applications of the VR-based intervention. Expanding the participant pool and including individuals with clinically diagnosed mysophobia would enhance the clinical relevance of the study. Additional work could also involve refining the game mechanics to include varying levels of contamination stimuli and adapting the game for different VR platforms to ensure accessibility. Exploring long term effects through repeated exposure and measuring changes in mysophobia symptoms over time would provide insights into the intervention's therapeutic potential. Finally, integrating biofeedback mechanisms into the game could enable real-time emotional regulation training. This would enhance the intervention's effectiveness further.


\section*{Conclusion}
This study demonstrates the feasibility of using a VR based gamified intervention to simulate contamination related scenarios and evaluate their impact on emotional states. By combining immersive VR technology with gamification, the study provides a controlled yet engaging platform for assessing responses to a common mysophobia trigger—sneezing. The results highlight the potential of VR based interventions for addressing contamination related anxieties to offer a safe and customizable environment for exposure therapy. However, limitations such as the small non mysophobic sample size and reliance on self-reported measures warrant further investigation. Future research should focus on scaling up the study with mysophobic patients and exploring the intervention's long-term therapeutic benefits. With continued refinement, VR based solutions have the potential to become valuable tools in the treatment of mysophobia and related anxiety disorders.

\bibliographystyle{plain}
\bibliography{reference.bib}

@article{lindner2020experiences,
  title={Experiences of gamified and automated virtual reality exposure therapy for spider phobia: qualitative study},
  author={Lindner, Philip and Rozental, Alexander and Jurell, Alice and Reuterski{\"o}ld, Lena and Andersson, Gerhard and Hamilton, William and Miloff, Alexander and Carlbring, Per and others},
  journal={JMIR serious games},
  volume={8},
  number={2},
  pages={e17807},
  year={2020},
  publisher={JMIR Publications Inc., Toronto, Canada}
}

@article{lindner2020gamified,
  title={Gamified, automated virtual reality exposure therapy for fear of spiders: a single-subject trial under simulated real-world conditions},
  author={Lindner, Philip and Miloff, Alexander and Bergman, Camilla and Andersson, Gerhard and Hamilton, William and Carlbring, Per},
  journal={Frontiers in psychiatry},
  volume={11},
  pages={116},
  year={2020},
  publisher={Frontiers Media SA}
}

@article{miloff2016single,
  title={Single-session gamified virtual reality exposure therapy for spider phobia vs. traditional exposure therapy: study protocol for a randomized controlled non-inferiority trial},
  author={Miloff, Alexander and Lindner, Philip and Hamilton, William and Reuterski{\"o}ld, Lena and Andersson, Gerhard and Carlbring, Per},
  journal={Trials},
  volume={17},
  pages={1--8},
  year={2016},
  publisher={Springer}
}

@article{davies2024immersive,
  title={Immersive and User-Adaptive Gamification in Cognitive Behavioural Therapy for Hypervigilance},
  author={Davies, Saskia and Owen, Tom and Walton, Sean P},
  journal={Authorea Preprints},
  year={2024},
  publisher={Authorea}
}

@article{kahlon2023gamified,
  title={Gamified virtual reality exposure therapy for adolescents with public speaking anxiety: a four-armed randomized controlled trial},
  author={Kahlon, Smiti and Lindner, Philip and Nordgreen, Tine},
  journal={Frontiers in Virtual Reality},
  volume={4},
  pages={1240778},
  year={2023},
  publisher={Frontiers Media SA}
}

@inproceedings{eishita2023gamified,
  title={Gamified Digital Intervention to Ameliorate the Aptitude of Exposure Therapy for OCD},
  author={Eishita, Farjana Z and Tasnim, Rifat Ara and Pongratz, Rick and Beard, David},
  booktitle={2023 IEEE Gaming, Entertainment, and Media Conference (GEM)},
  pages={1--4},
  year={2023},
  organization={IEEE}
}

@article{yuan2018application,
  title={Application of a Conscious System for a Study on Obsessive-compulsive Disorder},
  author={Yuan, Tianbai and Wang, Jianyu and Takeno, Junichi},
  journal={Procedia computer science},
  volume={145},
  pages={646--651},
  year={2018},
  publisher={Elsevier}
}

@article{qadir2019questionnaire,
  title={Questionnaire based study about association between blood oxygen level and Mysophobia},
  author={Qadir, Muhammad Imran and Yameen, Iqra Ali},
  journal={Biomed. J. Sci. Tech. Res},
  volume={14},
  pages={1--3},
  year={2019}
}

@misc{page2016virtual,
  title={Virtual reality exposure therapy for anxiety disorders: small samples and no controls?},
  author={Page, Sarah and Coxon, Matthew},
  journal={Frontiers in Psychology},
  volume={7},
  pages={326},
  year={2016},
  publisher={Frontiers Media SA}
}

@article{craske2015optimizing,
  title={Optimizing exposure therapy for anxiety disorders: an inhibitory learning and inhibitory regulation approach},
  author={Craske, Michelle},
  journal={Verhaltenstherapie},
  volume={25},
  number={2},
  pages={134--143},
  year={2015},
  publisher={S. Karger AG Basel, Switzerland}
}

@book{keedwell2015latin,
  title={Latin Squares and Their Applications: Latin Squares and Their Applications},
  author={Keedwell, A Donald and D{\'e}nes, J{\'o}zsef},
  year={2015},
  publisher={Elsevier}
}

@inproceedings{hossan2024adaptive,
  title={Adaptive Game Design Using Machine Learning Techniques: A Survey},
  author={Hossan, Md Mosharaf and Fouda, Mostafa M and Eishita, Farjana Z},
  booktitle={2024 IEEE International Conference on Internet of Things and Intelligence Systems (IoTaIS)},
  pages={144--150},
  year={2024},
  organization={IEEE}
}

@article{watson1988development,
  title={Development and validation of brief measures of positive and negative affect: the PANAS scales.},
  author={Watson, David and Clark, Lee Anna and Tellegen, Auke},
  journal={Journal of personality and social psychology},
  volume={54},
  number={6},
  pages={1063},
  year={1988},
  publisher={American Psychological Association}
}

@article{spielberger1971state,
  title={The state-trait anxiety inventory},
  author={Spielberger, Charles D and Gonzalez-Reigosa, Fernando and Martinez-Urrutia, Angel and Natalicio, Luiz FS and Natalicio, Diana S},
  journal={Revista Interamericana de Psicologia/Interamerican journal of psychology},
  volume={5},
  number={3 \& 4},
  year={1971}
}
\clearpage



\end{document}